\documentstyle{article}
\newcommand{\be}{\begin{equation}}
\newcommand{\ee}{\end{equation}}
\newcommand{\bea}{\begin{eqnarray}}
\newcommand{\eea}{\end{eqnarray}}
\newcommand{\bfm}{{\bf m}}
\newcommand{\tr}{{\rm tr}}
\newcommand{\pf}{{\rm pf}}
\newcommand{\dg}{^{\dagger}}
\newcommand{\nn}{\nonumber \\}
\begin{document}

\title{Orthogonality constraints and entropy in the SO(5)-Theory 
of High$T_c$-Superconductivity}

\author{Franz J. Wegner\\
Institut f\"ur Theoretische Physik\\
Ruprecht-Karls-Universit\"at Heidelberg, Germany}

\maketitle

\begin{abstract}
S.C. Zhang has put forward the idea that 
high-temperature-superconductors can be described in the 
framework of an SO(5)-symmetric theory in which the three 
components of the antiferromagnetic order-parameter and the two 
components of the two-particle condensate form a five-component 
order-parameter with SO(5) symmetry. Interactions small in 
comparison to this strong interaction introduce anisotropies into
the SO(5)-space and determine whether it is favorable for the 
system to be superconducting or antiferromagnetic.
Here the view is expressed that Zhang's derivation of the 
effective interaction $V_{\rm eff}$ based on his Hamiltonian 
$H_{\rm a}$ is not correct.
However, the orthogonality constraints introduced several pages 
after this 'derivation' give the key to an effective interaction 
very similar to that given by Zhang. It is shown that the 
orthogonality constraints are not rigorous constraints, but they 
maximize the entropy at finite temperature.
If the interaction drives the ground-state to the largest 
possible eigenvalues of the operators under consideration 
(antiferromagnetic ordering, superconducting condensate, etc.), 
then the orthogonality constraints are obeyed by the 
ground-state, too.

Key words: Superconductivity; Strongly correlated electrons

\end{abstract}

\section{Introduction}

In a recent paper \cite{Zhang} Zhang has proposed the idea that 
the components of the electron-pair condensate and those of the 
antiferromagnetic order-parameter form a five-component order 
parameter in an approximately SO(5)-invariant theory. 
Anisotropies small in comparison to the SO(5)-invariant 
interaction break this symmetry. They depend in particular on the 
chemical potential and thus on doping. This allows him to desribe 
the transition from antiferromagnetism in the half-filled system 
to superconductivity under moderately weak doping.

This phenomenological picture has been tested by numerical 
calculations for small Hubbard-systems \cite{Hubbard,tJ-Hubbard} 
and $t$-$J$-models \cite{tJ,tJ-Hubbard}, for which good agreement 
with the predictions from Zhang's theory are obtained. 
It has also been shown that spin-ladders observe SO(5)-symmetry 
\cite{ladder1,ladder2} or even SO(8)-symmetry \cite{SO8}.

Returning to higher-dimensional systems, it is pointed out here 
that Zhang's derivation of the effective interaction 
$V_{\rm eff}$ based on his Hamiltonian $H_{\rm a}$ seems to be 
not correct.
(Actually we struggled for quite a while trying to understand his 
derivation).
However, the orthogonality constraint introduced several pages 
after this 'derivation' gives the key to an effective interaction 
very similar to that given by Zhang. After having established 
this in sect. 2 it is shown in sect. 3, that the entropy contains 
squares of these constraint terms. 
Thus the orthogonality constraints are not strict requirements, 
but their fulfillment maximizes the entropy and lowers the free 
energy. 
Finally in sect. 4 it is argued, that also in the ground-state 
these constraints are likely to hold.

\section{Effective Interaction and the Orthogonality Constraint}
 
Zhang introduces a five-component order-parameter $n_i$, 
$i=1..5$. 
The operators associated with these components are given 
explicitely in the appendix.
Two components, $n_1$ and $n_5$, describe the real part and 
the imaginary part of the superconducting condensate, the 
three other components $n_2$, $n_3$, and $n_4$, are the three 
Cartesian components of the staggered magnetization. 
He assumes that a strong interaction which may be described by 
a Ginzburg-Landau interaction, leads to a symmetry breaking 
below a critical temperature $T_c$ and has an SO(5) symmetry, 
thus allowing both superconducting and antiferromagnetic 
order. 
This leading interaction is a function of
$\sum_i n_i^2$ only and does not give any preference to 
superconducting or antiferromagnetic order.

In addition there is a weaker interaction $H_{\rm a}$ which 
introduces anisotropy into the system
\be
H_{\rm a} = \frac{L^2_{1,5}}{2\chi_c} + 
\frac{L^2_{1,2}+L^2_{1,3}+L^2_{1,4}+L^2_{2,5}+L^2_{3,5}
+L^2_{4,5}}{2\chi_{\pi}} 
+ \frac{L^2_{2,3}+L^2_{2,4}+L^2_{3,4}}{2\chi_s} 
-\frac g2(n^2_2+n^2_3+n^2_4) -2\mu L_{1,5}.
\ee
The $L_{a,b}$ are operators bilinear in the electron creation 
and annihilation operators. They are given explicitely in 
appendix A. $\mu$ is the chemical 
potential. The operators $L$ obey $L_{b,a}=-L_{a,b}$ and the
commutator relations
\be
[L_{a,b},L_{c,d}]=-i\delta_{b,c}L_{a,d}+i\delta_{a,c}L_{b,d}
+i\delta_{b,d}L_{a,c}-i\delta_{a,d}L_{b,c} \label{comm}
\ee
of the orthogonal group SO(5), moreover the vector $n$ is 
rotated by $L$,
\be
[L_{a,b},n_c]=-i\delta_{b,c}n_a+i\delta_{a,c}n_c.
\ee
The derivative terms of Zhang's
interaction are left out here, since only the global state is 
investigated.
The effective potential is determined as the minimum of the 
Hamiltonian
for given order-parameter $n_1$, ... $n_5$, which Zhang 
normalizes to
\be
\sum_i n_i^2 = 1. \label{norm}
\ee
Since one considers the interaction
for a macroscopic system, all the components of $L$ and $n$ 
can be considered
as classical quantities and the obvious minimum is obtained 
for $L_{1,5}=2\mu\chi_c$, 
whereas all other components of $L$ vanish, which
yields what I call $V^{\rm naiv}_{\rm eff}$
\be
V^{\rm naiv}_{\rm eff}(n) = -2\mu^2 \chi_c - \frac g2 
(n^2_2+n^2_3+n^2_4).
\ee
It would mean that the only effect of the chemical potential 
is a lowering of the energy of the system, but it would have 
no effect on an anisotropy in the order-parameter space. 
This obviously is at variance with Zhang's claim
for the effective interaction
\be
V^{\rm Zhang}_{\rm eff}(n) = -2\mu^2(n^2_1+n^2_5)
[\chi_c(n^2_1+n^2_5)+\chi_{\pi}(n^2_2+n^2_3+n^2_4)]
-\frac g2 (n^2_2+n^2_3+n^2_4).
\ee
An indication that Zhang's Lagrangian
\bea
{\cal L}_{\rm a}&=&\sum_{a<b} \frac{\chi_{ab}}2 \omega_{ab}^2 
+ V(n) \\
\omega_{ab}&=&n_a(\partial_{\tau}n_b-iB_{bc}n_c)
-(a\rightarrow b) \label{omega}
\eea
is not equivalent to $H_{\rm a}$, is obvious from 
the fact that in Zhang's Lagrangian the only allowed rotations 
are in the 
plain spanned by $n$ and $\partial_{\tau}n - iBn$, whereas it 
does not
permit a rotation around a second perpendicular plain (which 
exists in 5 
dimensions) as it should. Thus the description by 
${\cal L}_{\rm a}$ is 
incomplete.

However, several pages later he introduces the constraints
\be
\epsilon_{a,b,c,d,e}n_cL_{d,e}=0. \label{ortho}
\ee
These constraints, (although not proven in Zhang's paper, 
since (\ref{omega}) 
is not granted) are the key for the effective interaction. 
If they are 
fulfilled, then one can conclude, that $L$ is of the form
\be
L_{a,b} = c_a n_b -c_b n_a. \label{cn}
\ee
(This can be found in the following way: If $n_1=1$, and all 
other $n_i$ 
vanish, then one finds from (\ref{ortho}) $L_{a,b}=0$, 
if both $a\not= 1$ and $b\not= 1$. 
However $L_{a,1}=-L_{1,a}=c_a$ for $a\not= 1$ can be chosen 
with arbitrary $c_a$. Thus one obtains in this special case 
eq. (\ref{cn}). The relation (\ref{cn}) is form-invariant 
under SO(5)-rotations. Thus it holds for general $n_i$.)
Eq. (\ref{cn}) has two consequences:

(i) One can now parametrize $H_{\rm a}$ in terms of the
$c_i$ and determine the minimum as one varies the coefficients 
$c$. One finds the minimum for
\be
c_2=c_3=c_4=0, \quad c_1=\frac{2\mu n_5}N, \quad 
c_5=-\frac{2\mu n_1}N
\ee
with
\be
N= \frac{n_1^2+n_5^2}{\chi_c} 
+ \frac{n^2_2+n^2_3+n^2_4}{\chi_{\pi}}.
\ee
{From} this one obtains the effective interaction
\be
V_{\rm eff} = -2\mu^2 \frac{n_1^2+n_5^2}N -\frac g2 
(n_2^2+n_3^2+n_4^2).
\ee
For $\chi_c=\chi_{\pi}$ it agrees with 
$V^{\rm Zhang}_{\rm eff}$.
If $\chi_c \not= \chi_{\pi}$, then the effective potentials are
different (actually in first order in $\chi_c-\chi_{\pi}$ 
there is still agreement), but many conclusions Zhang has 
drawn will continue to hold.

(ii) The constraints imply, that out of the representations 
$(\lambda_1,\lambda_2)$ the only allowed states have 
representations of the form
$(\lambda_1,0)$. Let me explain this shortly. 
The irreducible representations
of SO(5) are characterized by two (non negative) numbers 
$(\lambda_1,\lambda_2)$
with $\lambda_1 \geq \lambda_2 \geq 0$. 
The first number is the largest 
eigenvalue of one of the operators $L$, say $L_{1,2}$. 
Next consider the subspace of the states in this 
representation with this eigenvalue $\lambda_1$ for $L_{1,2}$. 
Then $\lambda_2$ is the largest eigenvalue
of one of the operators $L$, that commutes with the first one,
which may be $L_{3,4}$ in this subspace.
Semiclassically, that is for large $\lambda_1$ and 
$\lambda_2$, one may
replace the operators $L_{a,b}$ by their expectation values. 
Then the
eigenvalues of the antisymmetric matrix $L$ are 
$\pm i\lambda_1$, 
$\pm i\lambda_2$ and 0. If $L$ has the form (\ref{cn}), then
\be
\lambda_1 = \sqrt{\sum_i c_i^2 \sum_j n_j^2 
- (\sum_i c_i n_i)^2}, \quad \lambda_2=0.
\ee

Therefore the main problem left is to understand why the 
orthogonality 
constraints should hold. This will be done in the next section.

\section{Entropy as Source of the Orthogonality Constraint}

We claim that the orthogonality constraints are an effect of 
entropy.
It will be shown that a system even without any interaction 
has an entropic parts of the form
\be
-\big(\epsilon_{a,b,c,d,e}n_cL_{d,e}\big)^2
\ee
which make it favorable for the system to obey the 
orthogonality constraints.
Thus (\ref{ortho}) is not a strict constraint, but if it is 
fulfilled, then the entropy assumes a maximum. 

\subsection{Entropy of the Heisenberg Antiferromagnet}

First consider the Heisenberg antiferromagnet as an example,
which Zhang has rightly mentioned in his paper. The Heisenberg
antiferromagnet consists of two sublattices with magnetizations
$\bfm_1$ and $\bfm_2$. We expand the entropy in powers of 
$\bfm_1^2$ and $\bfm_2^2$. 
Since the two sublattice-magnetizations describe the behaviour
on different sublattices, the entropy is a sum of 
contributions on these sublattices
\be
S=S_0-c_1(\bfm_1^2+\bfm_2^2)-c_2\big((\bfm_1^2)^2
+(\bfm_2^2)^2\big)-...
\ee
Then denoting the homogeneous magnetization by $\bfm_0$ and 
the staggered magnetization by $\bfm_{q_0}$ we may write 
$\bfm_1=\bfm_0+\bfm_{q_0}$ and $\bfm_2=\bfm_0-\bfm_{q_0}$ 
and obtain the entropy
\be
S=S_0-2c_1(\bfm_0^2+\bfm_{q_0}^2) 
-2c_2\big(\bfm_0^2+\bfm_{q_0}^2\big)^2
-8c_2\big(\bfm_0\cdot\bfm_{q_0}\big)^2-...
\ee
Thus if there is no coupling in the interaction between 
$\bfm_0$ and $\bfm_{q_0}$, the system prefers to have the 
homogeneous and the staggered magnetization orthogonal to each 
other due to the contribution
$-8c_2(\bfm_0\cdot\bfm_{q_0})^2$ in the entropy. 
Remember that the entropy enters
into the free energy with a minus sign ($F=E-TS$), and thus
$\bfm_0\cdot\bfm_{q_0}=0$ will yield the minimum of the free 
energy.
Thus there is not a strict constraint 
$\bfm_0\cdot\bfm_{q_0}=0$ for the Heisenberg antiferromagnet, 
but there is a
term proportional to the square $(\bfm_0\cdot\bfm_{q_0})^2$
in the free energy, which favors the constraint to be obeyed. 

\subsection{The entropy}

It will now be shown that the entropy of the more general 
SO(5)-invariant system contains terms of type 
$(\epsilon_{a,b,c,d,e}n_cL_{d,e})^2$.
We will actually enlarge the system of operators $L_{a,b}$ to 
those of an SO(8) (for details see in the appendix). 
To determine the entropy we start from the Hamiltonian
\be
H_{\Omega} = \sum_{a,b} \Omega_{a,b} L_{a,b}, \quad 
\Omega_{b,a}=-\Omega_{a,b}. \label{HOmega}
\ee
where the set of our $L_{a,b}$ also includes the 
$n_a=L_{a,0}$. 
We do no longer use the normalization (\ref{norm}). 
We note, that recently a U(4)-scheme including the operators 
$L_{a,b}$ with $a,b=0..5$ and $L_{6,7}$ and their 
sub-groups has been considered in \cite{U4} (note that SU(4) 
is isomorphous to SO(6)). 
The $\Omega$ are introduced as Lagrange multipliers and will 
be adjusted to yield given expectation values of $L_{a,b}$, 
and the entropy will be calculated up to fourth order in $L$.

$\Omega$ is an eight-dimensional antisymmetric real matrix. 
Since the matrix is antihermitean, its eigenvalues
are purely imaginary and occur pairwise. 
Thus the eigenvalue equation can be written
\be
\sum_b \Omega_{a,b} (x^{(k)}_b \pm i y^{(k)}_b) 
= \pm i \omega^{(k)} (x^{(k)}_a \pm i y^{(k)}_a)
\ee
with $k=1..4$ and real vectors $x^{(k)}$ and $y^{(k)}$. 
The vectors $x^{(k)}$ and $y^{(k)}$ are orthogonal to each 
other. 
If they are normalized, then $\Omega$ may be represented
\be
\Omega_{a,b} = \sum_k \omega^{(k)} (x^{(k)}_a y^{(k)}_b 
- x^{(k)}_b y^{(k)}_a).
\ee
Next we perform a special orthogonal transformation, so 
that $x^{(k)}$ and $y^{(k)}$ are oriented in appropriate 
directions, e.g. so that $x^{(1)}$, $y^{(1)}$ point in the 
5- and 1- direction, $x^{(2)}$, $y^{(2)}$ in the 2- and 
3-direction, $x^{(3)}$ and $y^{(3)}$ in the 0- and 
7-direction and $x^{(4)}$ and $y^{(4)}$ in the 4- and 
6- direction, resp. 
After this special orthogonal transformation $H$ reads
\bea
H_{\rm trans} &=& 2(\omega^{(1)} N^0_{0,1,1} 
+ \omega^{(2)} N^0_{0,1,\sigma^z}
+\omega^{(3)} N^0_{0,g,1} -\omega^{(4)} N^0_{0,g,\sigma^z}) \nn
&=& \sum_{k,s} ( \omega^{(1)} +\omega^{(2)} s 
+ \omega^{(3)} g(k) -\omega^{(4)} g(k) s) (c\dg_{k,s} c_{k,s} 
- \frac 12). \label{Htrans}
\eea
(with $s=\pm 1$, $g(k)=\pm 1$, $g(k+q_0)=-g(k)$). 
Thus for momenta $k$ and $k+q_0$ one has in total $2^4$
states constructed out of four single-particle states, which 
may be either occupied or unoccupied and which contribute the 
energies $\pm \frac 12 \epsilon_{s,g}$ depending on whether 
the state is occupied or unoccupied
\be
\epsilon_{s,g} = \omega^{(1)} + \omega^{(2)} s 
+ \omega^{(3)} g - \omega^{(4)} gs.
\ee
Then we obtain for the partition function $Z$
\be
\ln Z=\sum \ln(\exp(\frac 12\beta\epsilon)
+\exp(-\frac 12\beta\epsilon))
=\sum (\ln 2 + \frac{x^2}2 - \frac{x^4}{12} + O(x^6)), 
\quad x=\frac{\beta\epsilon}2.
\ee
Summation over the four states yields
\bea
\sum x^2 &=& \beta^2 \sum_i \omega^{(i)2} \\
\sum x^4 &=& \frac{\beta^4}4 (\sum_i \omega^{(i)4} 
+6 \sum_{i<j} \omega^{(i)2} \omega^{(j)2} 
- 24 \omega^{(1)} \omega^{(2)} \omega^{(3)} \omega^{(4)} ).
\eea

Similarly we expand the entropy $S$
\be
S=k_B \sum (\ln 2 - \frac{x^2}2 + \frac{x^4}4 + O(x^6))
\ee
and determine the expectation value of the quantity $L^{(i)}$ 
conjugate to $\omega^{(i)}$ (only the first component is 
given; the others are obtained by permutation),
\be
L^{(1)} = \frac{\partial \ln Z}{2\beta\partial \omega^{(1)}} 
= \frac{\beta}2 \omega^{(1)}
-\frac{\beta^3}{24}\big(\omega^{(1)3} 
+ 3\omega^{(1)}(\omega^{(2)2}+\omega^{(3)2}+\omega^{(4)2})
-6\omega^{(2)}\omega^{(3)}\omega^{(4)} \big) + O(\omega^5)
\ee
and express $\omega^{(i)}$ in terms of $L^{(i)}$,
\be
\beta\omega^{(1)} = 2L^{(1)} + \frac 23\big( L^{(1)3}
+3L^{(1)}(L^{(2)2}+L^{(3)2}+L^{(4)2})
-6L^{(2)}L^{(3)}L^{(4)} \big) +O(L^5).
\ee
In diagonal form one has $L^{(1)}=L_{5,1}$, etc.
Then $S$ for the subspace of the electrons with momenta $k$ 
and $k+q_0$ reads
\be
S/k_B=4\ln2-2\sum_i L^{(i)2} -\frac 13 \big(\sum_i L^{(i)4} 
+6\sum_{i<j} L^{(i)2} L^{(j)2}
-24L^{(1)}L^{(2)}L^{(3)}L^{(4)}\big) +O(L^6). \label{SLeig}
\ee
Thus we have expressed the entropy in terms of the eigenvalues 
$\pm i L^{(i)}$ of the $8\times8$-matrix $(L_{a,b})$ of the 
expectation values $L_{a,b}$. 
We will express it now by the matrix-elements of $L$.
First we have
\bea
\sum_i L^{(i)2} &=& -\frac 12 \tr L^2 = -\sum_{a<b} L_{a,b}^2, 
\\
\sum_i L^{(i)4} &=& \frac 12 \tr L^4 = \sum_{a<b} L_{a,b}^4 
+ 2 \sum_{a,b<c} L_{a,b}^2 L_{a,c}^2 
+ 4\hspace{-5mm} \sum_{a<b,a<c<d,b\not=c,b\not=d} 
\hspace{-11mm} 
L_{a,c}L_{a,d}L_{b,c}L_{b,d}, \\
\hspace{-2mm} \sum_i L^{(i)4} + 2\sum_{i<j} L^{(i)2} L^{(j)2} 
&=& \frac 14 (\tr L^2)^2 
= \sum_{a<b} L_{a,b}^4 + 2\sum_{a,b<c} L_{a,b}^2 L_{a,c}^2 
+2\hspace{-3mm} \sum_{a<b,a<c<d,b\not=c,b\not=d} \hspace{-6mm}
L_{a,b}^2 L_{c,d}^2, \\
L^{(1)}L^{(2)}L^{(3)}L^{(4)} &=& \pf(L),
\eea
where $\pf(L)$ is the Pfaffian of $L$. 
In our case the indices $a$, $b$ of the elements $L_{a,b}$ are 
numbered from 0 to 7. 
Then the Pfaffian (which is defined only for antisymmetric 
matrices) is the sum $\sum_{k=1}^7 (-)^{k-1} L_{0,k} P_k$, 
where $P_k$ is the Pfaffian of the matrix obtained from the 
matrix $L$ by deleting the rows and columns with index 0 (that 
is the first one) and index $k$. 
One continues recursively until the Pfaffian of a matrix with 
no entry is left, which is defined to equal 1. 
(If one starts with a matrix of odd dimension, then
finally one arrives at the Pfaffian of the $1\times1$ matrix 
with entry 0, since it has to be antisymmetric. 
This Pfaffian is defined to equal 0. 
Therefore the Pfaffian of a matrix of odd dimensions vanishes.)
We mention that the determinant of an antisymmetric matrix 
equals the square of its Pfaffian. 
For $2\times2$ and $4\times4$ matrices the Pfaffians read
\bea
\pf\left(\begin{array}{cc} 0 & L_{a,b} \\ -L_{a,b} & 0 
\end{array} \right) &=& L_{a,b} \\
p_{a,b,c,d}:= \pf \left(\begin{array}{cccc} 
0 & L_{a,b} & L_{a,c} & L_{a,d} \\
-L_{a,b} & 0 & L_{b,c} & L_{b,d} \\
-L_{a,c} & -L_{b,c} & 0 & L_{c,d} \\
-L_{a,d} & -L_{b,d} & -L_{c,d} & 0
\end{array}\right) &=& L_{a,b} L_{c,d} - L_{a,c} L_{b,d} 
+ L_{a,d} L_{b,c}.
\eea
Then we have
\be
S/k_B=4\ln 2 + \tr L^2 -\frac 13\big(\frac 34 (\tr L^2)^2
-\tr L^4 -24 \pf(L) \big).
\ee
(Note that $\tr L^2$ is negative). 
Further algebraic manipulations yield
\bea
-\tr L^4 &=& -\frac 12 (\tr L^2)^2 
+ 4\sum_{a<b<c<d} p_{a,b,c,d}^2, \\
24\pf(L) &=& 4\sum_{a<b<c<d,\, e<f<g<h} 
\epsilon_{a,b,c,d,e,f,g,h} 
\,p_{a,b,c,d} \,p_{e,f,g,h}.
\eea 
Putting all the contributions together we obtain
\bea
S/k_B &=& 4\ln2 + \tr L^2 -\frac 1{12} (\tr L^2)^2 \nonumber \\
&-&\frac 43 \sum_{0<a<b<c} \big( p_{0,a,b,c} 
- \sum_{d<e<f<g} \epsilon_{a,b,c,d,e,f,g} \, p_{d,e,f,g} 
\big)^2 + O(L^6).
\eea
This corresponds to the separation of (\ref{SLeig}) into 
\bea
\hspace{-5mm}S/k_B &=& 4 \ln 2 -2 \sum_i L^{(i)2} 
- \frac 13(\sum_i L^{(i)2})^2 \nonumber \\
&-&\frac 43 \big( (L^{(1)}L^{(2)}-L^{(3)}L^{(4)})^2 
+ (L^{(1)}L^{(3)}-L^{(2)}L^{(4)})^2 
+ (L^{(1)}L^{(4)}-L^{(2)}L^{(3)})^2 \big) +O(L^6). 
\label{SLeig2}
\eea
Thus the entropy consists of a completely rotational invariant 
contribution depending only on $\tr L^2$ and a negative sum of 
squares of $p_{0,a,b,c} \pm p_{d,e,f,g}$. 
Note that the sum over $d,e,f,g$ contains exactly one term 
$\pm p$.
Thus a maximum of the entropy is reached, when the arguments 
of all the squares vanish. 
Provided $L_{a,b}$ vanishes if $a$ or $b$ equals 6 or 7 as 
assumed in the SO(5) theory, then out of the 35 squares 20 
vanish identically, 10 have the form 
$\epsilon_{a,b,c,d,e} n_c L_{d,e}$, which when required to
vanish are Zhang's orthogonality constraints, and 5 have the 
form $\epsilon_{a,b,c,d,e}L_{b,c}L_{d,e}$. 
One can easily see from (\ref{cn}), that if Zhang's 
orthogonality constraints are fulfilled, then also these
latter quantities vanish. 
Thus the orthogonality constraints are not a strict
requirements, but their fulfillment lowers the free energy.

We mention, that the vanishing of the squares in the second 
line of (\ref{SLeig2}) implies two types of solutions, either
one $L^{(i)}$ can be different from zero and the other 
$L^{(j)}$ vanish, or all $|L^{(i)}|$ are equal, and the 
product of the four $L^{(i)}$ is positive.

At half-filling (no doping) one has $L_{1,5}=0$. 
Then in the antiferromagnetic state, that is for non-vanishing 
components $L_{2,0}$, $L_{3,0}$, and $L_{4,0}$, but otherwise 
vanishing components $L$, only one eigenvalue $L^{(i)}$ is 
different from zero.
As soon as one has some doping $L_{1,5}$ is different from 0. 
If the system is still antiferromagnetic, then more than one 
eigenvalue differs from zero. 
On the other hand, if the modes associated with $n_6=L_{6,0}$ 
or $n_7=L_{7,0}$ are massive, as assumed, then it costs energy 
to have all four eigenvalues $L^{(i)}$ different from zero. 
Then it becomes preferable to have the superconductiong 
components $n_1=L_{1,0}$ and $n_5=L_{5,0}$ different from zero 
and to have vanishing antiferromagnetism, since then
again one has only one nonvanishing eigenvalue $L^{(i)}$.

\section{Limits for the ground state}

The entropy argument given in the preceeding section can be 
applied in the vicinity of the critical temperature. 
At low temperatures the contribution of the entropy to the 
free energy decreases, since it enters with the factor $T$.
Therefore we consider separately the situation at low 
temperatures.

\subsection{Antiferromagnet}

Let us start with a simple consideration for the 
antiferromagnet. 
Assume again two sublattices with magnetization $\bfm_1$ and 
$\bfm_2$. 
Assume they are restricted by upper bounds $|\bfm_1|\le 1$, 
$|\bfm_2|\le1$.
Suppose now the system has a homogeneous magnetization 
$\bfm_0$. 
If now a staggered magnetization parallel to $\bfm_0$ is 
added, then apparently $|\bfm_{q_0}|\le 1-|\bfm_0|$. 
If however the staggered magnetization is perpendicular to the 
homogeneous one, then one has the weaker restriction
$|\bfm_{q_0}|\le \sqrt{1-\bfm_0^2}$. 
Thus the antiferromagnetic interaction can act more strongly, 
if the staggered magnetization is perpendicular to the
homogeneous one. 
Thus again $\bfm_0\cdot\bfm_{q_0}=0$ is fulfilled.

\subsection{Hartree-Fock-Bogoliubov ground state}

We have seen, that at low temperatures the bounds on the 
appropriate quantities (order-parameters) are important for 
the ordering of the ground-state.
Therefore we will consider the bounds of the eigenvalues 
$\pm i L^{(i)}$ of the matrix $L$.
If for a fixed matrix $\Omega$ in (\ref{HOmega}) one takes the 
low temperature limit $\beta\rightarrow\infty$, then normally 
a pure Hartree-Fock-Bogoliubov state instead of a mixed state 
remains. 
Depending on the sign of $\epsilon_{s,g}$ the occupation 
number $n_{s,g}$ assumes one of its extremal values 0 and 1 
in the diagonal representation $H_{\rm trans}$ (\ref{Htrans})
\be
n_{s,g} = \frac 12 -\frac 12 {\rm sign} \epsilon_{s,g}.
\ee
{From} this we obtain the eigenvalues $L^{(i)}$
\bea
L^{(1)} &=& \frac 12 \sum_{s,g} n_{s,g} - 1 \\
L^{(2)} &=& \frac 12 \sum_{s,g} s\, n_{s,g} \\
L^{(3)} &=& \frac 12 \sum_{s,g} g\, n_{s,g} \\
L^{(4)} &=& \frac 12 \sum_{s,g} sg\, n_{s,g}.
\eea
Putting now $n_{s,g}=0$ or 1 in all combinations one finds two 
types of solutions for $L^{(i)}$. 
In the first class one $L^{(i)}=\pm 1$ and the other 
$L^{(j)}=0$, whereas in the second class one has 
$L^{(i)}=\pm \frac 12$ for all $i$ with the restriction for 
the signs, that the product of all $L^{(i)}$ is positive. 
One easily realizes that in all these cases the squares in the
second line of eq. (\ref{SLeig2}) vanish. 
As a consequence
the orthogonality constraints are fulfilled for these states. 
The eigenvalues obtained for the Hartree-Fock-Bogoliubov 
ground-state are the extremal ones. 
For correlated states these eigenvalues can only be reduced.
More precisely, the range of the set of possible eigenvalues 
$L^{(1)}$, ... $L^{(4)}$ lies in the convex volume bounded by 
the extremes given above.
For an interaction quadratic in the operators
$L$ the energy assumes extremal eigenvalues for extremal 
eigenvalues $L$ in the case of symmetry breaking. 
Without symmetry breaking the eigenvalues 
would vanish. 
We have argued before in favour of only one non-vanishing 
eigenvalue $L^{(i)}$. 
Reduced to SO(5) this would be $\lambda_1$, whereas
the second largest (vanishing one) is $\lambda_2$ of the 
representation $(\lambda_1,\lambda_2)$. 
These correspond to the representations ($\lambda_1$,0) 
actually found in the numerical calculations
\cite{tJ,tJ-Hubbard}.

\section{Conclusion}

We have shown that the orthogonality constraints introduced by 
Zhang in his SO(5)-theory of high$T_c$-superconductivity play 
an important role in the mechanism for the transition from 
antiferromagnetism to superconductivity as a function of 
doping. 
We have further shown that these constraints are
not strictly fulfilled, but that their fulfillment yields a 
maximum of the entropy (for fixed $\sum L_{a,b}^2$). 
At low temperatures the entropy plays a weaker role. 
However, if the interaction drives the ground-state to the 
case of extremal eigenvalues $L^{(i)}$, which may be very well 
the case for an interaction bilinear in the operators $L$, 
then again the orthogonality constraints are fulfilled.

We have expanded our scheme to a (mathematically natural) 
SO(8)-scheme.
In this scheme two types of solutions for maximal entropy or 
extremal eigenvalues appear. 
Since the added degrees of freedom are probably massive,
only those solutions, for which one eigenvalue is different 
from zero and the other ones vanish yield the minimum in the 
free energy.

These considerations did not take the microscopic interaction
seriously into account. 
Work in this direction has to be done.

{\bf Acknowledgment} I am indebted to Andreas Mielke and 
J\"urgen Stein for useful comments.

\begin{appendix}

\section{Operators in an SO(8) space}

\subsection{Bilinear operators}

All the operators in the SO(5) theory are of the form
\bea
N^+_{q,f,\gamma}&=&\frac 12 \sum_{k,s,t} f(k) 
c\dg_{k+q,s} (\gamma \sigma^y)_{s,t} c\dg_{-k,t}, \\
N^0_{q,f,\gamma}&=&\frac 12 \sum_{k,s,t} f(k) \gamma_{s,t}
(c\dg_{k+q,s} c_{k,t} - \frac 12\delta_{q,0}\delta_{s,t}), \\
N^-_{q,f,\gamma}&=&\frac 12 \sum_{k,s,t} f(k)
c_{k+q,s} (\sigma^y \gamma)_{s,t} c_{-k,t}.
\eea
Here the summation $k$ runs over the Brillouin zone. 
The vector $q$ can either be 0 or $q_0$, where $2q_0$ is a 
reciprocal lattice vector. The staggered magnetization is 
described by the wave-vector $q_0$ as above.
The function $f(k)$ stands either for 1 or $g(k)$ with 
$g(k)=g(-k)=-g(k+q_0)=\pm 1$.
Finally $\gamma$ is a hermitean two by two matrix. It may be 
either the unit matrix or one of the Pauli matrices.

First we consider the symmetry of $N^+$. If we exchange the 
two $c\dg$-operators, then we obtain
\be
N^+_{q,f,\gamma} = N^+_{q,f(.+q),\sigma^y\gamma^T\sigma^y}.
\ee
One has
\be
f(.+q)=s_{q,f}f \mbox{ with } 
s_{q,f}=\left\{ \begin{array}{c} 1 \mbox{ for } q=0 
\mbox{ or } f=1, \\
-1 \mbox{ for } q=q_0 \mbox{ and } f=g. \end{array} \right.
\ee
and
\be
\sigma^y\gamma^T\sigma^y = s_{\gamma} \gamma
\mbox{ with } s_{\gamma} = \left\{ \begin{array}{c} +1 
\mbox{ for } \gamma=1 \\
-1 \mbox{ for Pauli matrices } \end{array} \right.
\ee
{From} this we conclude that only the six operators 
$N^+_{0,1,1}$, $N^+_{0,g,1}$, $N^+_{q_0,1,1}$, 
$N^+_{q_0,g,\sigma_\alpha}$ with $s_{q,f}s_{\gamma}=1$
are different from zero. The same holds for $N^-$.

The hermitean adjoint operators are
\bea
(N^-_{q,f,\gamma})\dg &=& s_{q,f} N^+_{q,f,\gamma\dg}, \\
(N^0_{q,f,\gamma})\dg &=& s_{q,f} N^0_{q,f,\gamma\dg}.
\eea

We obtain for the commutators
\bea
{[}N^+_{q,f,\gamma},N^+_{q',f',\gamma'}]&=&0, \\
{[}N^-_{q,f,\gamma},N^-_{q',f',\gamma'}]&=&0, \\
{[}N^-_{q,f,\gamma},N^+_{q',f',\gamma'}]
&=&-2N^0_{q'-q,ff'(.-q),\gamma\gamma'}, \\
{[}N^-_{q,f,\gamma},N^0_{q',f',\gamma'}]
&=&N^-_{q-q',f(.-q')f',\gamma\gamma'}, \\
{[}N^0_{q,f,\gamma},N^+_{q',f',\gamma'}]
&=&N^+_{q+q',f(.+q')f',\gamma\gamma'}, \\
{[}N^0_{q,f,\gamma},N^0_{q',f',\gamma'}]
&=&\frac 12(N^0_{q+q',f(.+q')f',\gamma\gamma'}
-N^0_{q+q',ff'(.+q),\gamma'\gamma}).
\eea

\subsection{Zhang's Operators}

The operators introduced by Zhang, which obey the commutator 
relations (\ref{comm}) of an SO-group were
\bea
L_{5,1}=&Q&=N^0_{0,1,1}, \\
L_{1+\alpha,1}=\frac 12 (\pi\dg_{\alpha}+\pi_{\alpha}), &&
\pi\dg_{\alpha}=N^+_{q_0,g,\sigma^{\alpha}}, \\
L_{5,1+\alpha}=\frac {-i}2 (\pi\dg_{\alpha}-\pi_{\alpha}), &&
\pi_{\alpha}=-N^-_{q_0,g,\sigma^{\alpha}}, \\
L_{1+\alpha,1+\beta}
=\epsilon_{\alpha,\beta,\gamma}S_{0,\gamma}, &&
S_{0,\gamma}=N^0_{0,1,\sigma^{\gamma}}, \\
n_1=L_{1,0}=\frac 12(\Delta\dg+\Delta), && 
\Delta\dg=iN^+_{0,g,1},\\
n_5=L_{5,0}=\frac i2(\Delta\dg-\Delta), && 
\Delta=-iN^-_{0,g,1},\\
n_{1+\alpha}=L_{1+\alpha,0}&=&S_{q_0,\alpha}
=N^0_{q_0,1,\sigma^{\alpha}}.
\eea
Here we have added $n_a=L_{a,0}$ to the group, since they obey
the same commutator relations.

\subsection{Extension to operators obeying an SO(8) group} 

Obviously one can expand the range of operators $L_{a,b}$ by 
including the other operators $N$ introduced above so that any 
pairs of particles, of holes or particle-hole pairs with total 
momentum 0 or $q_0$ in the singlet and triplet channel appear. 
This allows us to introduce components $L_{a,b}$ with $a,b=6$ 
or 7. 
This expansion does not imply, that the symmetry group for the 
phase transition will be enlarged. 
Actually one expects that the new components of the 
'order-parameter' stay massive at the transition and will not 
contribute directly to the symmetry-breaking.

It is wellknown \cite{Lieb, Yang} that there are additional 
operators adding or removing two electrons, $\eta\dg$ and 
$\eta$, which commute with the Hubbard Hamiltonian, which 
allows us by means of the commutator relations (\ref{comm}) 
to introduce
\bea
L_{1,6}=\frac i2(\eta-\eta\dg), && \eta\dg=N^+_{q_0,1,1} \\
L_{5,6}=\frac 12(\eta\dg+\eta), && \eta=N^-_{q_0,1,1}, \\
L_{1+\alpha,6}&=&-N^0_{0,g,\sigma^{\alpha}}, \\
n_6=L_{6,0}&=&iN^0_{q_0,g,1}.
\eea

Finally there are 7 more operators left, which fulfill the 
appropriate
commutator relations with
\bea
L_{1,7}=\frac 12(\tilde{\eta}\dg+\tilde{\eta}), && 
\tilde{\eta}\dg=N^+_{0,1,1}, \\
L_{5,7}=\frac i2(\tilde{\eta}\dg-\tilde{\eta}), &&
\tilde{\eta}=N^-_{0,1,1}, \\
L_{1+\alpha,7} &=& iN^0_{q_0,g,\sigma^{\alpha}}, \\
L_{6,7} &=& -N^0_{q_0,1,1}, \\
n_7=L_{7,0} &=& -N^0_{0,g,1}.
\eea

\end{appendix}

\end{document}